\documentclass[preprint,12pt]{elsarticle}

\usepackage{lineno,hyperref}
\modulolinenumbers[1]

\journal{Information Processing Letters}

\usepackage{amsmath}
\usepackage{amsfonts}
\usepackage{framed}
\usepackage{amssymb}
\usepackage{natbib}
\usepackage{vaucanson-g}

\newtheorem{thm}{Theorem}
\newtheorem{lem}[thm]{Lemma}
\newcommand{\spc}{\hspace{.7pt}}
\newdefinition{definition}{Definition}
\newproof{proof}{Proof}
\newproof{proofofthm}{Proof of Theorem \ref{thm:PL-CF-1-Regular}}

\begin{document}

\begin{frontmatter}

\title{Regularity of languages generated by non context-free grammars over a~singleton terminal alphabet}

\author[address1,address2]{Alberto Pettorossi\corref{1}}
\cortext[1]{Corresponding author.}
\ead{pettorossi@info.uniroma2.it}

\author[address2]{Maurizio Proietti}
\ead{maurizio.proietti@iasi.cnr.it}
\address[address1]{DICII, University of Rome Tor Vergata, Rome, Italy}
\address[address2]{CNR-IASI, Rome, Italy}

\begin{abstract}
It is well-known that: (i)~every context-free language over a singleton 
terminal alphabet is 
regular~\cite{Har78}, and (ii)~the class of languages that satisfy 
the Pumping Lemma (for context-free languages)
is a proper super-class of 
the context-free languages. We show that 
any language in this super-class over a singleton terminal alphabet is regular. 
Our proof is based on an elementary transformational approach and 
does not rely on Parikh's 
Theorem~\cite{Par66}. Our result extends previously known results because 
there are languages that are not context-free, 
do satisfy the Pumping Lemma, 
and do not satisfy the hypotheses of Parikh's 
Theorem~\cite{Ra&98}. 

\end{abstract}

\begin{keyword}
Context-free languages, pumping lemma (for context-free languages), 
Parikh's Theorem, regular languages.
\end{keyword}

\end{frontmatter}


Let us begin by introducing our terminology and notations.

The set of the natural numbers
is denoted by $N$. The set of the \mbox{$n$-tuples} of natural numbers is denoted by $N^{n}$.
We say that a language $L$ is over the terminal alphabet $\Sigma$ iff $L\!\subseteq\!\Sigma^{*}$.
Given a word $w\!\in\!\Sigma^{*}$,
$w^{0}$ is the empty word~$\varepsilon$, 
and, for any $i\!\geq\!0$, $w^{i+1}$ 
is $w^{i}\,w$, that is, the concatenation of $w^{i}$ and $w$. 
The length of a word $w$ is denoted by $|w|$. 
Given a symbol $a\!\in\!\Sigma$,  the number of occurrences of~$a$ in $w$
is denoted by $|w|_{\displaystyle a}$. The cardinality of a set $A$ is denoted
by $|A|$.

Given an alphabet $\mathrm{\Sigma}$ such that $|\Sigma|\!=\!1$,
the concatenation of any two words $w_{1}, w_{2}$ in $\mathrm{\Sigma}^{*}$
is commutative,
that is, $w_{1}\,w_{2}=w_{2}\,w_{1}$.

\smallskip
In Theorem~\ref{thm:PL-CF-1-Regular} below we extend the well 
known result stating that any context-free language over a singleton terminal 
alphabet is a regular language~\cite{Har78}. 
An early proof of this result appears in a paper 
by Ginsburg and Rice~\cite{GiR62}. That proof 
is based on Tarski's fixpoint 
theorem and it is not based on the Pumping Lemma 
(contrary to what has been stated
in a paper by Andrei et al.~\cite{An&03}).
Our extension is due to the facts that: (i)~our proof does not rely on 
Parikh's Theorem~\cite{Par66}, like the proof in Harrison's book~\cite{Har78}, and
(ii)~there are non context-free languages that do
satisfy the Pumping Lemma 
(see Definition~\ref{def:pumpinglemma-cf})
and do not satisfy Parikh's Condition (see Definition~\ref{def:parikh-condition})
(and thus Parikh's Theorem cannot be applied)~\cite{Ra&98}. 
Our proof is very much related to one presented in a book 
by~Shallit~\cite{Sha08}, but we believe that ours is more elementary.

\begin{definition}[Pumping Lemma {\rm{\cite{Ba&61}}}]\label{def:pumpinglemma-cf}
{\rm{We}} {\rm{say that a language $L\!\subseteq\! \Sigma^{*}$  satisfies
the Pumping Lemma (for context-free languages) 
iff the following property, denoted  
$\mathit{PL}(L)$, holds:\\
$\exists\, n \!>\!0$,  $\forall\, z\in L$,
if $|z|\geq n$, then $\exists\, u,v,w,x,y\in \mathrm{\Sigma}^*$, such that

{\rm{(1)}}$~$ {\textup{$z=u\,v\,w\,x\,y$,}}

{\rm{(2)}}$~$ {\textup{$v\,x\not=\varepsilon$,}}

{\rm{(3)}}$~$ {\textup{$|v\,w\,x|\leq n$,  ~and }}

{\rm{(4)}}$~$ {\textup{$\forall\, i\geq0$, $u\,v^i\,w\,x^i\,y\in L$.}}
}}\hfill $\Box$
\end{definition}

\begin{definition}[Parikh's Condition~{\rm{\cite{Par66}}}]\label{def:parikh-condition}
{\rm{(i)~A subset $S$ of $N^{n}$ is said to be a {\em linear} set iff there exist
$v_{0},\ldots,v_{k}\!\in\!N^{n}$ such that
$S\!=\!\{v_{0}+n_{1}\,v_{1}+\ldots+n_{k}\,v_{k}\mid n_{1},\ldots,n_{k}\!\in\!N\}$}},
where, for any given $u\!=\!\langle u_{1},\ldots,u_{n}\rangle$ and 
$v\!=\!\langle v_{1},\ldots,v_{n}\rangle$ in 
$N^{n}$, $u\!+\!v$ denotes $\langle u_{1}\!+\!v_{1},\ldots,u_{n}\!+\!v_{n}\rangle$
and, for any \mbox{$m\!\geq\!0$}, $m\, u$ denotes $\langle m \,u_{1},\ldots,m \, u_{n}\rangle$.
(ii)~Given the alphabet $\Sigma\!=\!\{a_{1},\ldots,a_{n}\}$, we say that a language $L\!\subseteq\! \Sigma^{*}$  satisfies
Parikh's Condition iff
$\{\langle |w|_{\displaystyle a_{1}},\ldots, |w|_{\displaystyle a_{n}}\rangle \mid w\!\in\!L\}$ 
is a finite union of linear subsets of~$N^{n}$.\hfill $\Box$
\end{definition}


Let us first state and prove the following lemma whose proof is by transformation from 
Definition~\ref{def:pumpinglemma-cf}.

\begin{lem}\label{lem:PumpingLemma-CFL-Card1}
{\rm{For any language $L$ over a terminal alphabet $\mathrm{\Sigma}$ such that $|\Sigma|\!=\!1$, 
$\mathit{PL}(L)$ holds iff the following property, denoted $\mathit{PL}1(L)$, holds:\\
$\exists\, n \!>\!0$,  $\forall\, z\!\in\! L$, 
if $|z|\!\geq\! n$, then $\exists\,p\!\geq\!0,$ $q\!\geq\!0,$ $m\!\geq\!0$, such that

${\rm{(1.1)}}~$ $|z|=p+q$,

${\rm{(2.1)}}~$ $0\!<\!q\!\leq \!n$,

${\rm{(3.1)}}~$ 
$0\!<\!m\!+\!q \!\leq\! n$, ~and

${\rm{(4.1)}}~$ $\forall\,s\in \mathrm{\Sigma}^*,$ $\forall\,i\!\geq\!0$, if $|s|=p+i\,q$, then $s\in L$.}}
\end{lem}

\vspace{-4mm}
\begin{proof}
If $|\mathrm{\Sigma}|\!=\!1$, then commutativity of concatenation
implies that in~$\mathit{PL}(L)$
we can replace
 $v\,w\,x$ by $w\,v\,x$, and 
$u\,v^{i}\,w\,x^{i}\,y$ by $u\,y\,w\,(v\,x)^{i}$. 
Then, we can replace: $u\,y$ by $\widetilde u$,
$v\,x$ by $\widetilde{v}$, and $(\exists\,  u,\,v,\,w,\,x,\, y)$ by 
$(\exists\, \widetilde u,\,\widetilde v,\,w)$.
Thus, from $\mathit{PL}(L)$, we get:
\noindent
{\textup{$\exists\, n \!>\!0$,  $\forall\, z\in L$, 
if $|z|\!\geq\! n$, then $\exists\, \widetilde u,\widetilde v,w\in \mathrm{\Sigma}^*$, such that}} 

${\rm{(1')}}~$ {\textup{$z=\widetilde u\, w\, \widetilde v$,}}

${\rm{(2')}}~$ {\textup{$\widetilde v\not=\varepsilon$,}}

${\rm{(3')}}~$ {\textup{$|w\,\widetilde v|\leq n$, and}}

${\rm{(4')}}~$ {\textup{$\forall\, i\!\geq\!0$, $\widetilde u\,w\,\widetilde v\spc ^i\in L$.}}

\smallskip\noindent
Now if we take the lengths of the words and we denote 
$|\widetilde u\,w|$ by $p$, $|\widetilde v|$ by $q$, and  $|w|$ by $m$, we get:

\noindent
{\textup{$\exists\, n \!>\!0$,  $\forall\, z\in L$, 
if $|z|\!\geq\! n$, then $\exists\,p\!\geq\!0,$ $q\!\geq\!0,$ $m\!\geq\!0$, such that}}\nopagebreak

${\rm{(1'')}}~$ {\textup{$|z|=p+q$,}}

${\rm{(2'')}}~$ {\textup{$q\!>\!0$,}}

${\rm{(3'')}}~$ {\textup{$m\!+\!q \leq n$, ~and}}

${\rm{(4'')}}~$ {\textup{$\forall\,s\!\in\! \mathrm{\Sigma}^*,$ $\forall\,i\!\geq\!0$, if  $|s|=p+i\,q$, then $s\in L$.}}

\smallskip\noindent
For all $n\!>\!0$, $q\!>\!0$, and $m\!\geq\!0$, we have that $(q\!>\!0 \wedge m\!+\!q \!\leq\! n)$
iff $(0\!<\!q\!\leq \!n ~\wedge~ 0\!<\! m\!+\!q \!\leq\! n)$.
Thus, we get $\mathit{PL}1(L)$. \hfill $\Box$

\end{proof}
\vspace{-2mm}

We say that $\mathit{PL}1(L)$ {\it holds for} $b$ if $b$ is a witness of
the quantification `$\exists\, n \!>\!0$' in $\mathit{PL}1(L)$.
The following theorem states our main result.\vspace{-1mm}

\begin{thm}\label{thm:PL-CF-1-Regular}
{\textup{Let $L$ be any language over a terminal alphabet $\Sigma$ such that $|\Sigma|\!=\!1$.
If $\mathit{PL}(L)$ holds, then $L$ is a regular language.}}
\end{thm}
\vspace{-8mm}

\noindent
\begin{proof}  
Without loss of generality,
let us consider a language~$L$ over the 
terminal alphabet~$\{a\}$, such that $\mathit{PL}(L)$ holds. 
By Lemma~\ref{lem:PumpingLemma-CFL-Card1}, we have that 
$\mathit{PL}1(L)$ holds for some 
positive integer $b$.
Let us consider the following two disjoint languages whose union is $L$:

\smallskip
\noindent
(i)~$L_{\displaystyle{<\!b}} = \{ a^{k}  \mid  a^{k}\!\in\! L \wedge k\!<\!b\}$ ~~and~~
(ii)~$L_{\displaystyle{\geq\!b}} = \{ a^{k}  \mid a^{k}\!\in\! L \wedge k\!\geq\! b\}$.

\smallskip
\noindent
Now, $L_{\displaystyle{<\!b}}$ is a regular language, because it is finite.\rule{0mm}{3.3mm}
Since regular languages are closed\rule{0mm}{3.7mm} under finite union and 
intersection~\cite{HoU79}, in order to prove
that $L$ is regular, it is enough to prove, as we now do, that 


\smallskip
$L_{\displaystyle \geq\! b} = \bigcup \mathcal{S} ~\cap~ \{a^{i}\mid i\!\geq\!b\} $ \hfill $(\dagger 1)$~

\smallskip
\noindent
where: (i)~$\mathcal S$ is a set of languages which is\rule{0mm}{3.7mm} a subset of the 
following {\rm finite} set $\mathcal L$ of languages ($k, p_{h}, q_{0}, \ldots, q_{k}$ are
integers):

\smallskip
$\mathcal{L} = \{L^{\displaystyle{\langle p_h,q_0,\ldots,q_k\rangle}} \mid 
~ (0\!\leq \!k\!<\!b)   
~\wedge~ (0\!\leq\! p_{h}\! <\!b) ~\wedge~ $\vspace*{-1mm} \hfill $(\dagger 2)$~\nopagebreak

\vspace{1.2mm}
\hspace{44mm}$ (0\!<\! q_{0}\! \leq\!b) ~\wedge \ldots
\wedge~ (0\!<\! q_{k}\! \leq\!b) ~\wedge~ $\nopagebreak

\hspace{45mm}$q_{0},\ldots,q_{k}$ are all distinct\,$\}$ 

\vspace{1mm}
\noindent and (ii)~for all $k, p_{h}, q_{0}, \ldots, q_{k}$, 
the language:\smallskip

$L^{\displaystyle{\langle p_h,q_0,\ldots,q_k\rangle}}\! =\! 
\{a^{\displaystyle{\hspace{.5mm}p_h+i_0 \, q_0+\ldots+i_k\, q_k}} \mid 
i_0\!>\!0 ~\wedge \ldots \wedge~ i_k\!>\!0\}$\hfill $(\dagger 3)$~

\smallskip
\noindent is {\rm regular}.

Indeed, 
(i)~$\{a^{i}\mid i\!\geq\!b\}$ is regular,
\noindent 
{(ii)}~$\mathcal{L}$ 
is a {\rm{finite}} set of languages  because, for any $b$, there exists only 
a finite number of tuples 
$\langle p_h,q_0,\ldots,q_k\rangle$ satisfying all the conditions
stated inside the set expression $(\dagger 2)$, and
(iii)~the language $L^{\displaystyle{\langle p_h,q_0,\ldots,q_k\rangle}}$ 
is {\rm regular} because it is recognized by the following
nondeterministic\rule{0mm}{3.5mm} finite 
automaton with initial state~$A$ and final state~$B$: 

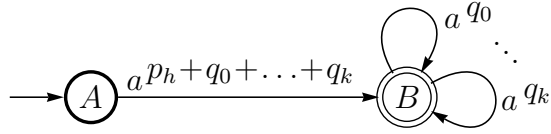
\hangindent=0mm

\vspace{-2mm}
\begin{center}
\VCDraw{%
\begin{VCPicture}{(0,-1.5)(10,2.5)}
\normalsize
\FixStateDiameter{12mm}
\FixStateLineDouble{0.4}{1.3}
\ChgStateLineWidth{1.2}
\SetEdgeArrowWidth{6pt}
\SetEdgeArrowLengthCoef{1.8}

\State[A]{(0,0)}{1}
\FinalState[B]{(7,0)}{2}
\SetStateLineColor{white}   
\FixStateDiameter{1mm}      
\State[.]{(9.0,1.2)}{dot1}  
\State[.]{(9.2,1.0)}{dot2}  
\State[.]{(9.4,0.8)}{dot3}  

\Initial{1}
\Edge{1}{2}\LabelL[0.47]{a^{\displaystyle{\hspace{.5mm}p_h\!+\!q_0\!+\!\ldots\!+\!q_k}}}
\LoopN{2}{}\LabelL[0.74]{\,a^{\displaystyle{\,q_0}}}
\LoopE{2}{}\LabelL[0.5]{a^{\displaystyle{\,q_k}}}
\end{VCPicture}}
\end{center}

\vspace{-4mm}

\noindent
In order to prove Equality~$(\dagger 1)$ it remains to prove that, for any 
$z\!\in\!L_{\displaystyle{\geq\!b}}$, there exists a tuple of the form
$\langle p_{h}, q_{0}, q_{1},\ldots, q_{k}\rangle$ such\rule{0mm}{3.7mm}  that 
$z\!\in\!L^{\displaystyle {\langle p_{h}, q_{0}, q_{1},\ldots, q_{k}\rangle}}$.

Given any word $z\!\in\!L_{\displaystyle{\geq\!b}}$, the following algorithm
constructs a tuple of the form $\langle p_{h}, q_{0}, q_{1},\ldots, q_{h}\rangle$,
for\rule{0mm}{3.7mm} some $h\!\geq\!0$.

\vspace{-2mm}
\begin{framed}
\vspace{-1mm}

\hspace{8mm}$\{\,z\!\in\!L_{\displaystyle{\geq\!b}}\,\}$ 
\hfill {\it Tuple Generation Algorithm}

\smallskip
\indent
$\ell\! :=\! |z|;~~~ i\!:=\!0;~~~ \langle p_{0},q_{0}\rangle\!:=\!\pi(\ell);$

\smallskip
\hspace{8mm}$\{\,|z|= p_{i} + \sum_{j=0}^{i} q_{j} ~~\wedge~~
\bigwedge_{j=0}^{i} ~0\!<\!q_{j}\!\leq\!b ~~\wedge~~ 0\!\leq\!p_{i}\,\} $

\smallskip
{\bf while}~ $p_{i}\!\geq\!b$ ~{\bf do}~ $\ell \!:=\! p_{i}; ~~ i \!:=\! i \!+\!1; ~~ 
\langle p_{i},q_{i}\rangle\!:=\!\pi(\ell)$ ~{\bf od};

\vspace{.7mm}
$h \!:=\! i;$ 

\hspace{8mm}$\{\,|z|= p_{h} + \sum_{j=0}^{h} q_{j} ~~\wedge~~
\bigwedge_{j=0}^{h} ~0\!<\!q_{j}\!\leq\!b ~~\wedge~~ 0\!\leq\! p_{h}\!<\!b\,\} \vspace*{-2mm}$

\end{framed}
\vspace*{-2mm}

\noindent
In this algorithm $\pi$ is a function from $N$ to $N\!\times\!N$, whose existence follows from the validity
of~$\mathit{PL}1(L)$, satisfying the following condition:
for every $\ell\!\geq\!b$, $\pi(\ell)\!=\!\langle p,q\rangle$ 
such that $\ell\! =\! p\!+\!q$ and $0\!<\!q\!\leq\!b$
(take $i\!=\!1$ in Condition~(4.1) of $\mathit{PL}1(L)$ in Lemma~\ref{lem:PumpingLemma-CFL-Card1}). 
The termination of the Tuple Generation 
Algorithm is a consequence of the fact that, for every~$z\!\in\! L_{\displaystyle{\geq\!b}}$, 
for every~$i\!\geq\!0$,  \mbox{$p_{i}\!=\!p_{i+1}+q_{i+1}$}
and $q_{i}\!>\!0$.
This implies that $p_{0},p_{1},\ldots$ is a\rule{0mm}{3.5mm} strictly decreasing sequence of integers,
and eventually in that sequence we will get an element smaller than $b$, and the while-loop 
terminates.

Thus, for every  $z\!\in\! L_{\displaystyle{\geq\!b}}$, 
there exist $h\!\geq\!0,$ $ p_0,$ $q_0,$ $p_1,$ $q_1,$ $p_2,$ $q_2,$ $\ldots,$ 
$p_h, q_h$  such that:

\smallskip
\makebox[6mm]{$z$}$=a^{\displaystyle{\,(p_h+q_h)+q_{h-1}+\ldots+q_2+q_1+q_0}}$\hfill$(\dagger 4)$~
\smallskip


\noindent
\makebox[12mm][l]{where:} 
\noindent
$0\!\leq\! p_h \!<\! b$ ~and~
for every $i$, if $0\!\leq\!i\!<\!h$, then
($p_{i}\!\geq \!b$ and
$0 \!<\! q_i\! \leq\! b$).

\smallskip
\noindent 
In general, in Equality~$(\dagger 4)$ the $q_i$'s
 are {\it{not\/}} all distinct.
Thus, by rearranging the summands, and writing $i_{j}\, q_{j}$, instead of  
$({q_{j}+\ldots+q_{j}})$ with $i_{j}$ occurrences of $q_{j}$, 
we have that,
for every word $z\!\in\! L_{\displaystyle{\geq\!b}}$, there exist
some integers $k,p_h,i_{0},q_0,\ldots,i_{k},q_k$ such that

\smallskip

$z=a^{\displaystyle{\,p_h+i_0\, q_0+\ldots+i_k \,q_k}}$,  ~~~ where:

\vspace{1mm}
\noindent
~{\makebox[9mm][l]{($\ell$\spc 0)}$0\!\leq\!k$,}\hspace{15mm} 
\noindent
~\makebox[9mm][l]{($\ell$\spc 1)}$0\!\leq\!p_h\!<\!b$,\hspace{15mm} 
\noindent
\noindent
~\makebox[9mm][l]{($\ell$\spc 2)}\makebox[51mm][l]{$0\!<q_0\!\leq\!b, \ldots, 0\!<q_k\!\leq\!b$,}

\noindent
~\makebox[9mm][l]{($\ell$\spc 3)}$q_0,\ldots,q_k$ are {all distinct}, ~~and~
~\makebox[9mm][l]{($\ell$\spc 4)}$i_0\!>\!0,\ldots, i_k\!>\!0$.
\
\vspace{1mm}

\noindent 
From ($\ell$\spc 2) and ($\ell$\spc 3), 
we have that $k\!<\!b$.
Hence, Condition~($\ell$\spc 0) can be strengthened to: 
\makebox[10mm][l]{($\ell\spc {0}^{*})$}$0\!\leq\!k\!<\!b$. We also have that $k\!\leq\!h$, and 
$k\!=\!h$ when in Equality~$(\dagger 4)$ the values of $q_{0},\ldots,q_{h}$ are all distinct.


\smallskip

Since Conditions~$(\ell\spc {0}^{*}$), $(\ell\spc {1})$,  
$(\ell\spc {2})$, and $(\ell\spc {3})$ are those occurring 
in the set expressions~$(\dagger 2)$, and Condition~$(\ell\spc {4})$ is the one 
occurring in the set expressions~$(\dagger 3)$, we have
concluded the proof of Equality~$(\dagger 1)$ and that of Theorem~\ref{thm:PL-CF-1-Regular}.~\hfill $\Box$
\end{proof} 

\smallskip
Let us make a few of remarks on the proof of Theorem~\ref{thm:PL-CF-1-Regular}.

\noindent
(i)~The validity of~$\mathit{PL}1(L)$ 
tells us that the function~$\pi$ exists, but it does not tell us
how to compute 
$\pi(\ell)$, for any given $\ell\!\geq\!b$.

\noindent
(ii)~Since summation is commutative, it may be the case that a language in~$\mathcal L$ corresponds to more than one tuple $\langle p_h,q_0,\ldots,q_k\rangle$. 
In particular, we have that $L^{\displaystyle{\langle p_h,q_0,\ldots,q_k\rangle}}$
$=$ $L^{\displaystyle{\langle p_h,q'_0,\ldots,q'_k\rangle}}$, whenever
$\langle q_0,\ldots,q_k\rangle$ is a permutation of
$\langle q'_0,\ldots,q'_k\rangle$.\rule{0mm}{3.5mm}   

\noindent
(iii)~If $b\!=\!1$, then $k\!=\!h\!=\!0$. Thus, from \mbox{Conditions~($\ell$\spc 1)} 
\mbox{and~($\ell$\spc 3)} we have: $\langle p_{h},q_{h}\rangle 
\!=\!\langle p_{0},q_{0}\rangle 
\!=\!\langle 0,1\rangle$. We also have that $\mathcal{L}$ is the singleton 
$\{L^{\textstyle{\langle 0,1\rangle}}\}$, where  
$L^{\textstyle{\langle 0,1\rangle}}$ is the language  $\{a^{i}\mid i\!>\!0\}$.

\noindent
(iv)~In Equality~$(\dagger 1)$ the set $\mathcal S$ of languages may be a {\it proper} subset of 
$\mathcal L$. Indeed, let us consider the language $L\!=\!\{a\,(a\,a)^{n}\mid n\!\geq\!0\}$
generated by the context-free grammar $S\rightarrow a\,S\,a \mid  a$.
Since $\mathit{PL}1(L)$ holds for $3$, we can take the constant $b$ 
occurring in Equality~$(\dagger 1)$ to be $3$.
If we consider the word $z\!=\!a\,a\,a$, 
then the set~$\mathcal L$ 
of languages includes, among others,  the languages
$L^{\langle 0,3\rangle}\!=\! \{a^{0\, +\, i \cdot 3}\mid i\!>\!0\}$, 
$L^{\langle 1,2\rangle}\!=\! \{a^{1\, +\, i \cdot 2}\mid i\!>\!0\}$, and 
$L^{\langle 2,1\rangle}\!=\! \{a^{2\, +\, i \cdot 1}\mid i\!>\!0\}$ (these three
languages are obtained for $k\!=\!h\!=\!0$).
Now, $L_{\displaystyle \geq 3}= L^{\langle 1,2 \rangle} \cap \{a^{i} \mid i\!\geq\!3\}=L^{\langle 1,2 \rangle}$, while $L^{\langle 0,3 \rangle}\!\not\subseteq 
\!L_{\displaystyle \geq 3}$ and $L^{\langle 2,1 \rangle}\!\not\subseteq 
\!L_{\displaystyle \geq 3}$.

\noindent
(v)~It may be the case that the\rule{0mm}{3.7mm} length ${p_{h}\!+\!q_{0}\!+\!\ldots\!+\!q_{k}}$ of the word labeling the arc 
from state $A$ to state $B$ of the finite automaton
depicted above, is smaller than $b$. Thus, in the definition of 
$L_{\displaystyle{\geq\!b}}$ the
intersection of  $\bigcup \mathcal S$ with $\{a^{i} \mid i\!\geq\!b\}$ 
ensures that only words whose length is\rule{0mm}{3.7mm} at least~$b$ are considered.




\section*{Acknowledgements}

This work has been partially funded
by INdAM-GNCS (Italy).


\end{document}